\documentclass[%
 aip,
 amsmath,amssymb,
 preprint,%
 reprint,%
]{revtex4-1}

\usepackage{graphicx}% Include figure files
\usepackage{dcolumn}% Align table columns on decimal point
\usepackage{bm}% bold math
\usepackage[utf8]{inputenc}
\usepackage[T1]{fontenc}
\usepackage{mathptmx}
\usepackage{etoolbox}

\makeatletter
\def\@email#1#2{%
 \endgroup
 \patchcmd{\titleblock@produce}
  {\frontmatter@RRAPformat}
  {\frontmatter@RRAPformat{\produce@RRAP{*#1\href{mailto:#2}{#2}}}\frontmatter@RRAPformat}
  {}{}
}%
\makeatother

\usepackage[nolist,nohyperlinks]{acronym} % INPUT OF ACROS AFTER \maketitel in revtex-4.1 !!!

\usepackage[hidelinks]{hyperref} 
\usepackage{orcidlink} % requires texLive2022 

\usepackage{multirow, array}
\usepackage{tabularray}

\begin{acronym}[ECU]

\acro{AWMPS}[AWMPS]{arbitrary waveform magnetic particle spectrometer}
\acro{ADC}[ADC]{analog-to-digital converter}
\acro{AUC}[AUC]{area under the curve}

\acro{CT}[CT]{computed tomography}

\acro{DAC}[DAC]{digital-to-analog converter}
\acro{DFG}[DFG]{drive-field generator}
\acro{DFCs}[DFCs]{drive-field coils}
\acro{DF}[DF]{drive-field}
\acro{DA}[DA]{differential amplifier}
\acro{DLS}[DLS]{dynamic light scattering}

\acro{ECD}[ECD]{equivalent circuit diagram}
\acro{ESR}[ESR]{equivalent series resistance}

\acro{FFL}[FFL]{field-free-line}
\acro{FFP}[FFP]{field-free-point}
\acro{FFR}[FFR]{field-free-region}
\acro{FFT}[FFT]{fast Fourier transform}
\acro{FOV}[FOV]{field-of-view}
\acro{FWHM}[FWHM]{full width at half maximum}

\acro{GUI}[GUI]{graphical user interface}

\acro{HCC}[HCC]{high current circuit}
\acro{HCR}[HCR]{high current resonator}

\acro{ICN}[ICN]{inductive coupling network}
\acro{ISI}[ISI]{integrated signal intensity}
\acro{ICs}[ICs]{integrated circuits}
\acro{IA}[IA]{instrumentation amplifier}

\acro{ICU}[ICU]{intensive care unit} 
\acro{LFR}[LFR]{low-field-region}
\acro{LNA}[LNA]{low noise amplifier}

\acro{MPI}[MPI]{magnetic particle imaging}
\acro{MRI}[MRI]{magnetic resonance imaging}
\acro{MTT}[MTT]{mean-transit-time}
\acro{MNPs}[MNPs]{magnetic nanoparticles}
\acro{MPS}[MPS]{magnetic particle spectroscopy}
\acro{MoM}[MoM]{method of moments}

\acro{PSF}[PSF]{point spread function}
\acro{PNS}[PNS]{peripheral nerve stimulation}
\acro{PTT}[PTT]{pulmonary transit time}

\acro{Q}[Q-factor]{quality factor}

\acro{RF}[RF]{radio-frequency}
\acro{RPs}[RPs]{RedPitaya STEMlab 125-14}
\acro{RF-fields}[RF]{radio frequency fields}
\acro{rBV}[rBV]{relative blood-volume}
\acro{rBF}[rBF]{relative blood-flow}
\acro{rCBV}[rCBV]{relative cerebral-blood-volume}
\acro{rCBF}[rCBF]{relative cerebral-blood-flow}
\acro{RBCs}[RBCs]{red blood cells}

\acro{SNR}[SNR]{signal-to-noise ratio}
\acro{SU}[SU]{surveillance unit}
\acro{SAR}[SAR]{specific absorption rate}
\acro{SPIONs}[SPIONs]{superparamagnetic iron-oxid nanoparticles}
\acro{USPIONs}[USPIONs]{ultrasmall superparamagnetic iron-oxid nanoparticles}
\acro{SF}{selection field}
\acro{SFG}[SFG]{selection-field generator}
\acro{SEM}[SEM]{scanning electron microscope}

\acro{TF}[TF]{transfer function}
\acro{THD}[THD]{total harmonic distortion}
\acro{TTP}[TTP]{time-to-peak}
\acro{TEM}[TEM]{transmission electron microscopy}

\acro{VOI}[VOI]{volume of interest}
\acro{VSM}[VSM]{vibrating sample magnetometer}

\end{acronym}

\usepackage{xcolor}
\definecolor{ibilight}{RGB}{193,216,237}
\definecolor{ibidark}{RGB}{0,73,146}	% UKE-blau
\definecolor{uke2}{RGB}{170,156,143} 	% Mocca
\definecolor{uke3}{RGB}{87,87,86}		% Schwarz (Schrift)
\definecolor{ukesec1}{RGB}{255,223,0}	% Gelb
\definecolor{ukesec2}{RGB}{239,123,5}	% Orange (dunkel)
\definecolor{ukesec3}{RGB}{104,195,205}	% "TUHH"
\definecolor{ukesec4}{RGB}{138,189,36}	% Grün
\definecolor{tuhh}{RGB}{45,198,214}     % TUHH

\newcommand{\mv}[1]{\boldsymbol{#1}}

\newcommand{\tu}[1]{\textup{#1}}

\usepackage[binary-units = true,exponent-product = \cdot]{siunitx}
\def\numx#1e#2#3{{#1}\cdot10^{#2#3}}
\DeclareSIUnit{\mup}{\text{$\mu_0$}}
\DeclareSIUnit{\sample}{S}
\DeclareSIUnit{\mT}{\milli\tesla}
\DeclareSIUnit{\uH}{\micro\henry}
\DeclareSIUnit{\proceduredefinedunit}{p.d.u.}

\begin{document}

\title[Resonant Inductive Coupling Network for Human-Sized MPI]{Resonant Inductive Coupling Network for Human-Sized\\ Magnetic Particle Imaging} %Title of paper

\author{Fabian Mohn\,\orcidlink{0000-0002-9151-9929}$^{1,2,\ast}$}%
\email[]{fabian.mohn@tuhh.de}
\author{Fynn F\"orger\,\orcidlink{0000-0002-3865-4603}$^{1,2}$} 
\author{Florian Thieben\,\orcidlink{0000-0002-2890-5288}$^{1,2}$} 
\author{Martin M\"oddel\,\orcidlink{0000-0002-4737-7863}$^{1,2}$}
\author{\\Ingo Schmale\,\orcidlink{0000-0003-1081-0085}$^{3}$}
\author{Tobias Knopp\,\orcidlink{0000-0002-1589-8517}$^{1,2,4}$} 
\author{Matthias Graeser\,\orcidlink{0000-0003-1472-5988}$^{4,5}$}
\noaffiliation

\affiliation{\footnotesize Institute for Biomedical Imaging, Hamburg University of Technology, 21073 Hamburg, Germany\looseness=-1}
\affiliation{\footnotesize Section for Biomedical Imaging, University Medical Center Hamburg-Eppendorf, 20251 Hamburg, Germany\looseness=-1}
\affiliation{\footnotesize Philips GmbH Innovative Technologies, Research Laboratories, 22335 Hamburg, Germany\looseness=-1}
\affiliation{\footnotesize Fraunhofer Research Institution for Individualized and Cell-based Medical Engineering, IMTE, 23562 Lübeck, Germany\looseness=-1}
\affiliation{\footnotesize Institute of Medical Engineering, University of Lübeck, 23562 Lübeck, Germany\looseness=-1}

\date{\today}

\begin{abstract} 
In Magnetic Particle Imaging, a field-free region is maneuvered throughout the field of view using a time-varying magnetic field known as the drive-field. Human-sized systems operate the drive-field in the kHz range and generate it by utilizing strong currents that can rise to the kA range within a coil called the drive field generator. Matching and tuning between a power amplifier, a band-pass filter and the drive-field generator is required. Here, for reasons of safety in future human scanners, a symmetrical topology and a transformer, called inductive coupling network is used.
Our primary objectives are to achieve floating potentials to ensure patient safety, attaining high linearity and high gain for the resonant transformer.
We present a novel systematic approach to the design of a loss-optimized resonant toroid with a D-shaped cross section, employing segmentation to adjust the inductance-to-resistance ratio while maintaining a constant quality factor. Simultaneously, we derive a specific matching condition of a symmetric transmit-receive circuit for magnetic particle imaging. The chosen setup filters the fundamental frequency and allows simultaneous signal transmission and reception. 
In addition, the decoupling of multiple drive field channels is discussed and the primary side of the transformer is evaluated for maximum coupling and minimum stray field. 
Two prototypes were constructed, measured, decoupled, and compared to the derived theory and to method-of-moment based simulations.
\end{abstract}

\pacs{}% insert suggested PACS numbers in braces on next line

\maketitle %\maketitle must follow title, authors, abstract and \pacs

%%%Author: place Acronyms here to work, AFTER maketitle and within \document

%%%%%%%%%%%%%%%%%%%%%%%%%%%%%%%%%%%%%%%%%%%%%%%%%%%%%%%%%%%%%%%%%%%%%%%%%%%%%%%%%%%%%%%%%%%%%%%%%%
%%%%%%%%%%%%%%%%%%%%%%%%%%%%%%%%%%%%%%%%%%%%%%%%%%%%%%%%%%%%%%%%%%%%%%%%%%%%%%%%%%%%%%%%%%%%%%%%%%
\section{Introduction}
\label{sec:intro}

Medical imaging modalities such as \ac{MPI} and \ac{MRI} rely on alternating current to generate strong radio-frequency fields that form the backbone of signal generation and acquisition~\cite{gleich_tomographic_2005,lauterbur_image_1973}. 
In the context of \ac{MPI}, the frequency of the so called drive field does not depend on the Larmor precession of hydrogen atoms, nor is it correlated to a static $B_0$ field, as in \ac{MRI}. The choice of this frequency is flexible, with the proviso that it should be above the human audible range, but below frequencies where wave propagation effects begin to affect signal detection and below the limits for energy deployment due to the tissue's \ac{SAR}~\cite{schmale_mpi_2015, dossel_safety_2013}. Typically, this frequency falls in the range of \SIrange{20}{160}{\kHz}~\cite{gleich_experimental_2008, schmale_human_2013,schmale_mpi_2015}, where lower frequencies tend to cause more \ac{PNS}~\cite{weinberg_increasing_2012}. Also, the best non-linear signal response of the required \ac{MNPs} that provide the image contrast in \ac{MPI} falls into this range, depending on the particle's anisotropy~\cite{tay_optimization_2020}. 
Currents for a human torso system reach the \si{\kilo\ampere} range~\cite{schmale_design_2015,sattel_setup_2015}, whereas head-sized system require around \SIrange{300}{500}{\A}~\cite{thieben_system_2024}.
While it is possible to perform \ac{MPI} with non-sinusoidal excitation waveforms~\cite{tay_pulsed_2019, mohn_system_2022}, the benefit of using sinusoidal excitation lies in the ability to implement resonators like passive filters.
This study focuses primarily on \ac{MPI}, while the basic concept has broader applicability to similar circuits and other frequencies. Such circuits can be found in the context of inductively coupled wireless power transmission~\cite{valtchev_electromagnetic_2012,morita_power_2017,mirbozorgi_smart_2014,yang_wireless_2020}, power converters~\cite{tong_3-d_2019, ziegler_air-core_2020}, 
band-pass filters~\cite{stelzner_toroidal_2015, bernacki_disturbance_2019, wang_design_2018, mattingly_drive_2022}, or other applications requiring high linearity and large currents.
A prominent challenge in this context is the formulation of a resonant transformer, i.e. the primary side forms a resonance with the band-pass filter output stage and the secondary side of the transformer is part of a high quality factor ($Q$) resonant transmit-receive circuit, called the \ac{HCR}.

%%%-------------------------------------------
%%% OVERVIEW: network system schematic %%%
\begin{figure*}[t!]
    \centering
    \includegraphics[width=0.9\linewidth]{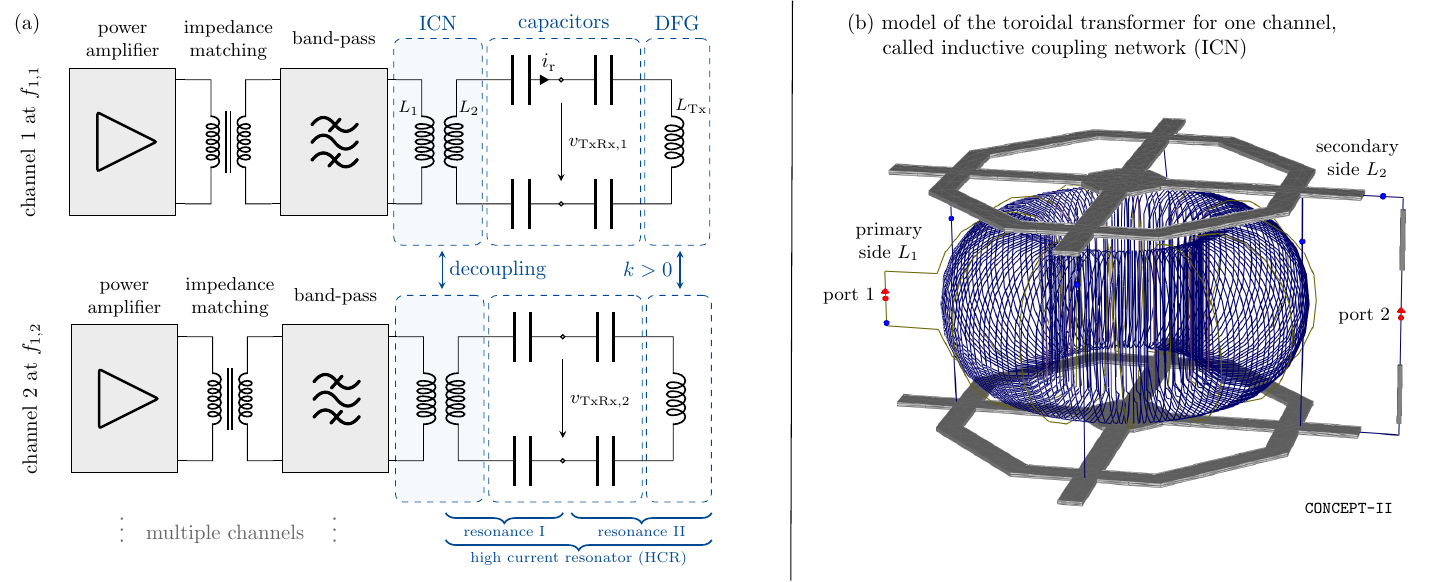}
    \caption{\label{fig:system_overview}\textbf{Overview of the MPI transmit chain and the ICN}. In (a), the power amplifier is connected to a ferrite-core transformer, a band-pass filter and to the primary side of the \ac{ICN}. The secondary side of this transformer is part of a high-Q resonator, called \acf{HCR}, that includes the \acf{DFG}. Magnetic particles excited by the \ac{DFG} induce signals via $L_\tu{Tx}$ and the receive signal is tapped within the symmetric \ac{HCR} at $v_\textup{TxRx}$. In (b), a 4-fold segmented toroidal, D-shaped \ac{ICN} is shown within a simulation setup of CONCEPT-II~\cite{institut_fur_theoretische_elektrotechnik_concept-ii_2023}.}
\end{figure*}
%%%-----------------------------------------

Our purpose of the resonant transformer, referred to as \acf{ICN}, is to obtain a differential voltage system and a safe operating voltage.
Ground loops and long, high-current cables can introduce undesired harmonics or disturbances into the receive spectrum, i.e., induced by eddy currents in nonlinear components such as unsoldered joints (screws). These distortions can dominate important particle harmonics or sidebands that are essential to the imaging process~\cite{paysen_characterization_2020, sattel_setup_2015}. 
A differential design with balanced filters avoids a global ground node and is less susceptible to interference and noise~\cite{johnson_filtering_1966}. Another strong advantage of a differential setup are floating signals with respect to a patient under examination, who is always capacitively coupled to ground. A single-ended transmit chain entails the danger for humans if they come into contact with any single point in the system. As \ac{MPI} systems strive for human trials~\cite{thieben_system_2024, vogel_impi_2023,schmale_human_2013,borgert_perspectives_2013}, 
this safety aspect can not be ignored. Furthermore, reducing voltage levels in proximity to the patient requires a low inductance \ac{DFG}, which in turn requires large currents to maintain the same field specifications~\cite{schmale_design_2015,thieben_system_2024}.

In this work, we design and implement a resonant toroidal inductive coupling network that encompasses concerns regarding safety and incorporates a strictly linear and loss-optimized design. We elaborate on our design decisions and weigh different conditions and restrictions to present a new approach of finding suitable parameter choices, i.e., for inductors, cross-section shape, parallel segments, gain and dimensions. Our reasoning is intended to be transferable to other transformers under similar constraints using resonant loads for applications beyond medical imaging.
Further, we consider crosstalk by channel decoupling between multiple drive-field channels that each use an individual \ac{ICN} and review multiple decoupling strategies. 
Based on a TxRx topology that was presented without details by Sattel et al.~\cite{sattel_setup_2015}, this study elaborates on the original implementation and optimization of an \ac{ICN} and tailors the design parameters to a human-sized head scanner~\cite{thieben_system_2024}.

%%%%%%%%%%%%%%%%%%%%%%%%%%%%%%%%%%%%%%%%%%%%%%%%%%%%%%%%%%%%%%%%%%%%%%%%%%%%%%%%%%%%%%%%%%%%
%%%%%%%%%%%%%%%%%%%%%%%%%%%%%%%%%%%%%%%%%%%%%%%%%%%%%%%%%%%%%%%%%%%%%%%%%%%%%%%%%%%%%%%%%%%%
\section{Motivation and Purpose}
\label{sec:motivation-purpose}

Many current \ac{MPI} systems use dedicated receive coils, often in a gradiometer configuration~\cite{graeser_analog_2013,graeser_towards_2017,radermacher_highly_2023,mcdonough_implementation_2022}, effectively suppressing feedthrough harmonics, interference and systemic background~\cite{paysen_improved_2018,paysen_characterization_2020}.
However, receive coils take up valuable space and ultimately increase power consumption when signal generating coils have to be placed farther away, independent the method of feedthrough suppression~\cite{borgert_perspectives_2013}.
For preclinical systems with rodent-sized bores, power consumption can be managed, however, on the path to human-sized systems, this issue becomes a major challenge~\cite{schmale_design_2015,borgert_perspectives_2013}.
The advantage of choosing to combine the transmit and receive circuitry is that it reduces system complexity and overall space requirements by eliminating the need for a dedicated receive coil nested within the \ac{DFG} and the receive band-stop filter~\cite{weizenecker_three-dimensional_2009}.
One possible topology of a transmit-receive (TxRx) circuit is a symmetric \ac{HCR} as shown in \autoref{fig:system_overview}\,(a) with a linear \ac{ICN}, shown in \autoref{fig:system_overview}\,(b). A comprehensive characterization of different noise and background contributions of TxRx and dedicated receive systems can be found in Ref.~\cite{paysen_characterization_2020}. However, there is currently no clear consensus in the community as to which system performs better on a human scale with manageable power consumption, although this may change in the future.
The development of a TxRx system with an \ac{ICN} is driven by various design goals, including a. achieving floating potentials, b. ensuring linearity, c. attaining high current gain and high $Q$, d. designing circuit symmetry for excitation, filtering and signal reception, and e. modularizing for impedance matching.

\noindent\paragraph{Floating potentials}
The first design goal concerns patient safety and avoids ground loops that may negatively affect signal reception. A galvanically isolated \acf{HCR} with an overall low voltage and floating potentials is achieved by many common transformer topologies. Floating potentials become relevant for a patient being examined in the scanner. Following the principle of the first fault case, the potential separation ensures that a patient is not exposed to a life-threatening current through single contact with the circuit. This would require contact at two separate points of the circuit, which drastically reduces the risk. For this unlikely event, we take the further precaution of reducing the overall voltage in the \ac{DFG} and \ac{HCR} by selecting low-inductance components~\cite{thieben_system_2024,ozaslan_design_2020}.
A capacitive coupling network~\cite{zhaksylyk_analysis_2020} would not provide real isolation at the drive-field frequency, although DC currents are blocked.

%%%-------------------------------------------
%%% THD analysis of iron 
\begin{figure*}[t!]
    \centering
    \includegraphics[width=1.0\linewidth]{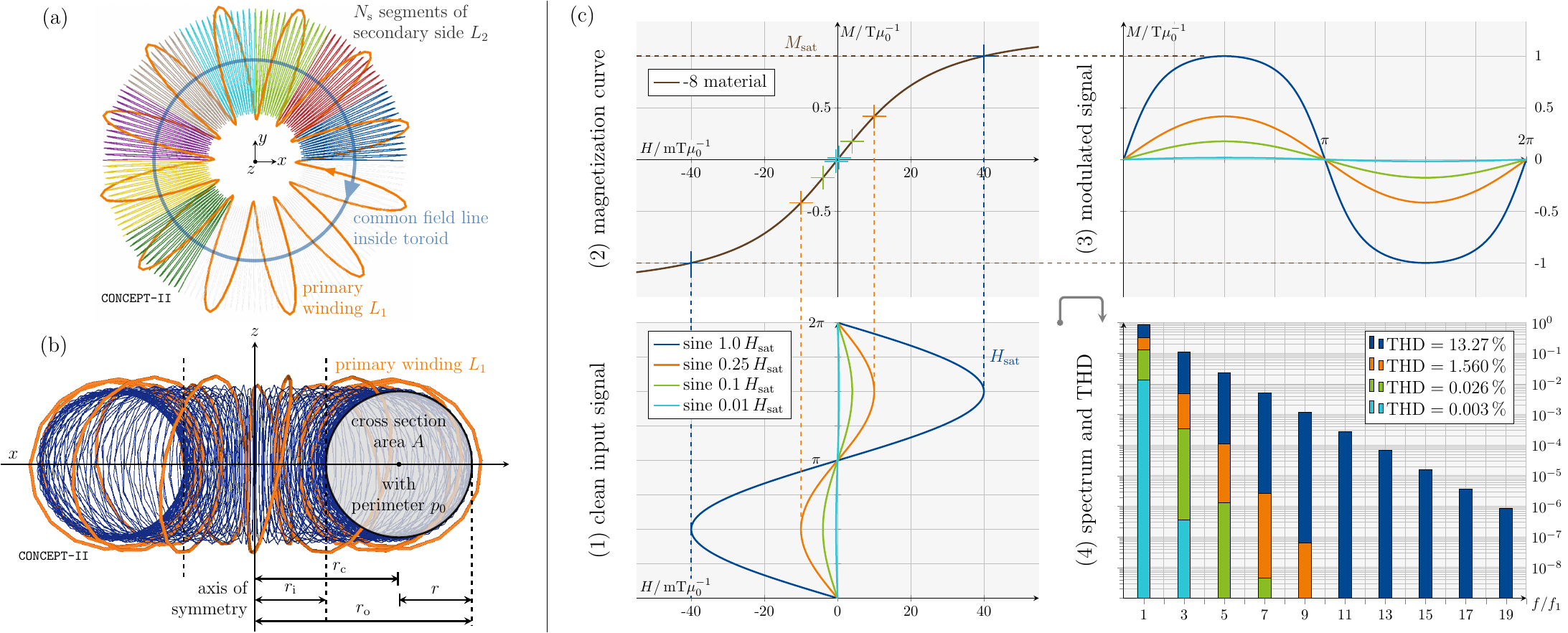}
    \caption{\label{fig:Seg+THD}\textbf{Segmented toroidal air-core transformer and THD simulation of iron-core saturation without hysteresis.}
    Shown in (a) is a toroid parallelized with $N_s=12$ individual segments, of which $8$ are shown in different colors ($4$ are hidden). The primary winding also forms a toroid, here with $N_1=12$ turns, whereas each secondary segment has $N_2=13$ turns. In (b), the circular cross-section in the $xz$-plane is shown with characteristic parameters noted.
    On the right in (c), three sinusoidal input signals (1) are modulated by the magnetization curve in (2), which results in the output in (3) and its spectrum in (4).
    The magnetization curve is adopted from the ``-8'' iron powder core material~\cite{micrometals_inc_datasheet_2007}, but without hysteresis effects and modeled with a $\tanh$-function~\cite{herceg_use_2020}.
    A THD of \SI{0.5}{\%}, typical for class-AB amplifiers, is caused by an amplitude of about \SI{0.14}{} of $H_\tu{sat}$.}
\end{figure*}
%%%-----------------------------------------

\noindent\paragraph{Linearity} 
A strictly linear air-core transformer is required, because the \ac{ICN} is located after the band-pass filter stage and must not increase the \ac{THD} of the transmit or the receive signal. Harmonics generated directly within the \ac{HCR}, where the receive signal is tapped, would be unfiltered and thus mask and alter the particle's response. 
Harmonic generation by iron-core transformers is well known, due to core saturation effects~\cite{daut_harmonic_2010,ramirez-nino_core_2016} and the principle is demonstrated in \autoref{fig:Seg+THD}\,(c). 
Here, we compare the \ac{THD} after the band-pass filter with the \ac{THD} generated by an ferrite-core transformer. 
Although an iron core would be preferable for its high permeability, effectively bundling magnetic field lines to achieve a high coupling coefficient $k$ and low leakage inductance, it introduces distortion due to saturation effects of the ferrite-core material.
This constraint is so strong, that a \ac{THD} of $0.5$\,\% from a class-AB power amplifier in combination with our transmit filter results in a theoretical \ac{THD} benchmark of below $5\cdot 10^{-6}$\,\%, calculated using the fundamental definition THD$_\textup{F}$ ~\cite{shmilovitz_definition_2005}.
The typical \ac{THD} of linear power amplifier ranges from $0.1\,\%$ to $1\,\%$~\cite{aetechron_datasheet_2023,aetechron_datasheet_2023-1,crown_datasheet_2023,drhubert_datasheet_2023} and our filter achieves a measured $65$\,dB amplitude attenuation at the second harmonic $f_2$, $100$\,dB at $f_3$ and below $150$\,dB at $f_4$ and above~\cite{mattingly_drive_2022, thieben_system_2024}.
Regarding a low-core loss iron powder material like ``-8'' (Micrometals, Inc., CA, USA~\cite{micrometals_inc_datasheet_2007}) and neglecting any hysteresis effects, a maximum amplitude deflection of only $1\,\%$ of the saturation field strength $H_\tu{sat}$ results in a \ac{THD} of $3\cdot 10^{-3}\,\%$, as shown in \autoref{fig:Seg+THD}\,(c)\,(4).
Although this number is extremely low, it is about 3 orders of magnitude higher than the combined \ac{THD} of the amplifier and band-pass filter. Likewise, an amplitude of about $0.14\cdot H_\tu{sat}$ would cause a \ac{THD} of $0.5$\,\%, counteracting any achievements by the band-pass filter. Moreover, intermodulation of harmonics will occur that further degrades the \ac{THD} of an iron-core transformer, which was not modeled here.
Consequently, an air-core transformer is preferred to achieve maximum linearity and a toroidal shape is advantageous to enclose the majority of field lines within the toroid. This minimizes any coupling leakage flux, eddy current losses in surrounding shielding and reduces the susceptibility to external disturbances to avoid spurious harmonics. %~\cite{hoult_principle_2000}. 

\noindent\paragraph{High current gain and high $Q$}
In order to maintain the same excitation field using low-inductance components, higher currents are required~\cite{thieben_system_2024}. We are focusing on minimal conduction losses and an overall loss-optimized transformer design.
The \ac{ICN} is a resonant transformer, shown in \autoref{fig:system_overview}\,(b), and the secondary side of the transformer is part of the \ac{HCR}. Note that both voltage and current are amplified on the secondary side, unlike non-resonant transformers. This is based on a two-step process of having a tank LC with high impedance from the perspective of the primary side and the voltage scaling into a high Q load on the secondary side. However, amplification is due to resonance and a change in the turns ratio still changes voltage and current in opposite directions. A limiting factor of the gain is the structural size, because it limits the overall achievable $Q$ of the \ac{HCR}. 

Independent of the desired inductor values, the cross-sectional shape of the toroid is optimized to maximize the inductance $L_2$ for a given wire length, i.e. the perimeter $p_0$ of the toroidal cross-section (D-shape). To achieve this, we implement the results of a comprehensive study by Murgatroyd~\cite{murgatroyd_optimum_1982}.
Additionally, frequency dependent losses are minimized by using litz wires that suppress the skin effect and reduce proximity effect losses~\cite{sullivan_optimal_1999}. 
Due to the angular symmetry of the toroid, the winding core offers the possibility of parallel segmentation of the secondary side. Such parallel segments share the same enclosed field lines to compose a single composite toroid as shown in \autoref{fig:Seg+THD}\,(a) and (b), which can be used to shift the desired value for $L_2$ at a constant $Q$ to change the desired turns ratio $n$.
A trade-off must be made between wire diameter, feasible parallelization of toroidal segments, stacked wire layers near the symmetry axis, heat dissipation, and target inductance.

\noindent\paragraph{TxRx circuit symmetry}
Symmetry refers to the \ac{HCR} layout, where the capacitors are split into two separate banks for each inductor, as shown in \autoref{fig:system_overview}\,(a). This offers the advantage of combining transmit and receive chains in a single circuit, as proposed by Sattel et. al~\cite{sattel_setup_2015}. Here, the \ac{HCR} acts simultaneously as a filter for the fundamental frequency, allowing to tap the particle's response $v_\textup{TxRx}$ during transmission. 
However, a partial attenuation of the particle signal is inevitable using this configuration because the strength of the particle signal relies on the proportion of inductors $L_2$ and $L_\tu{Tx}$, which we call the \ac{DFG} matching condition.

\noindent\paragraph{Impedance Matching Modularization} 
For feasible impedance matching, a modularization into two matching stages is carried out, as shown in \autoref{fig:system_overview}. The \ac{ICN} can thus be designed without additional constraints regarding amplifier load matching, which is done separately by an amplifier matching stage. To this end, we use an ferrite-core transformer before the band-pass filter with a maximum amplitude below 10\,\% of $H_\tu{sat}$, as explained in paragraph~b., on the condition that the transformer's \ac{THD} is similar to the amplifier's \ac{THD} and a band-pass filter is implemented afterwards.

%%%%%%%%%%%%%%%%%%%%%%%%%%%%%%%%%%%%%%%%%%%%%%%%%%%%%%%%%%%%%%%%%%%%%%%%%%%%%%%%%%%%%%%%%%%%
%%%%%%%%%%%%%%%%%%%%%%%%%%%%%%%%%%%%%%%%%%%%%%%%%%%%%%%%%%%%%%%%%%%%%%%%%%%%%%%%%%%%%%%%%%%
%%%%%%%%%%%%%%%%%%%%%%%%%%%%%%%%%%%%%%%%%%%%%%%%%%%%%%%%%%%%%%%%%%%%%%%%%%%%%%%%%%%%%%%%%%%
\section{Theory}
\label{sec:theory}

In this section, we present a systematic optimization of an \acl{ICN}, delineating the theory and criteria employed to derive the configurations for both sides of the toroidal transformer. Our approach commences with an examination of the transformer's current gain $G$ in \autoref{sec:theory:TFbasics} and \ref{sec:theory:volume}. The focus then shifts to optimizing the toroidal geometry with the goal of minimizing losses and maximizing the inductance as described in \autoref{sec:theory:topology}. This optimization process encompasses cross-sectional shape, multiple winding layers, segmentation, and primary winding characteristics.
In order to select an appropriate inductance value, we proceed to the inductance matching condition specific to a symmetric setup of the \ac{HCR} in \autoref{sec:theory:DFGmatch}, a consideration tailored for the \ac{MPI} imaging modality. 

Given the broader context of multichannel \ac{MPI} systems, it is imperative to devise individualized \ac{ICN}s for each channel and account for coupling, which is described in \autoref{sec:theory:decoupling}. Nevertheless, up to that point, the entire section focuses on a single channel with the resonance frequency $f_1$.

%%%%%%%%%%%%%%%%%%%%%%%%%%%%%%%%%%%%%%%%%%%%%%%%%%%%%%%%%%%%%%%%%%%%%%%%%%%%%%%%%%%%%%%%%%%%
%%%%%%%%%%%%%%%%%%%%%%%%%%%%%%%%%%%%%%%%%%%%%%%%%%%%%%%%%%%%%%%%%%%%%%%%%%%%%%%%%%%%%%%%%%%%
\subsection{Maximum Transformer Current Gain}
\label{sec:theory:TFbasics}

\begin{figure}[t!]
    \centering
    \includegraphics[width=1.0\linewidth]{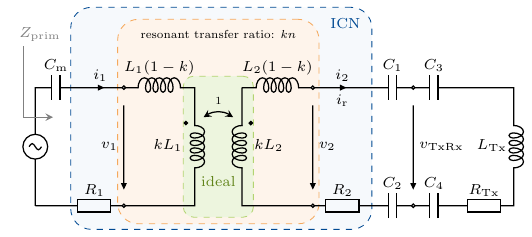}
    \caption{\label{fig:ECD}\textbf{Resonant transformer circuit diagram.} A resonant transformer is shown, where an ideal transformer remains embedded in the center. The inductors are split into leakage and coupled inductances. 
    The matching capacitor $C_\tu{m}$ resembles the filter output stage, which goes into resonance with the leakage inductance $L_{\sigma,1}=L_1(1-k)$ and $Z_\tu{prim}$ becomes real at $\omega=2\pi f_1$.
    }
\end{figure}

The current amplification factor, denoted as $G$ for the \ac{ICN} transformer, will be investigated with the objective to maximize it, according to design goal~c. 
A first expression for $G$ can be derived by considering the power $P_2$ on the secondary side, caused by the primary current $i_1$ that induces the voltage $v_2$ as in
\begin{equation}\label{eq:P2}
    P_2 = \frac{|v_2|^2}{R_\tu{s}} 
        = \frac{|j \omega M i_1|^2}{R_\tu{s}}
        = \frac{\omega^2 M^2 |i_1|^2}{R_\tu{s}} \; ,
\end{equation}
where $\omega = 2\pi f_1$ denotes the channel angular frequency, $j$ the imaginary unit, $M$ the mutual inductance of the transformer, and $R_\tu{s}=R_2+R_\tu{Tx}$ the total series resistance of the secondary side at resonance (capacitor losses are neglected)~\cite{mett_mri_2016}. All components, currents and voltages are shown in \autoref{fig:ECD}.
Now, the power at resonance can be rewritten to obtain the current gain $G$ by
\begin{equation}\label{eq:G2}
    |i_2|^2 R_\tu{s} = \frac{\omega^2 M^2 |i_1|^2}{R_\tu{s}}  ~~\Rightarrow ~~  G= \frac{i_2}{i_1} = \frac{\omega M}{R_\tu{s}} \; .
\end{equation}
In the following, this expression is rearranged for resonant transformers, using characteristic transformer variables like the turns ratio $n$, the coupling coefficient $k \in [0,1]$, and the quality factor $Q$ of the transformers secondary side~\cite{ayachit_transfer_2016, witulski_introduction_1995}.
Note that voltages across the leakage inductance $L_{\sigma,1}=L_1(1-k)$ and matching capacitor $C_\tu{m}$ of the band-pass filter cancel at resonance. Further, $L_2$ is part of a resonant circuit on the secondary side, therefore voltage and current are increased.

%%%-------------------------------------------
%%% ECD only %%%
%%%-------------------------------------------

The turns ratio is defined by the number of primary turns $N_1$ to secondary turns $N_2$. For an ideal transformer that is perfectly coupled ($k=1$), there is no phase difference and therefore no losses between the two sides. In this case, $n$ directly relates the magnitude of the voltages $v_1$ and $v_2$ and can be used to step up or step down the reflected impedance seen at the input~\cite{witulski_introduction_1995}, as given in
\begin{equation} \label{eq:n} 
    n = \dfrac{N_1}{N_2} = \sqrt{\dfrac{L_1}{L_2}} = \frac{v_1}{v_2}\; .
\end{equation}
$L_1$ and $L_2$ are the transformers primary and secondary inductances.
Next, to obtain the coupling coefficient, we first regard the coupling factors 
\begin{align} \label{eq:k12}
    K_{1} = \frac{\Phi_{21}}{\Phi_{11}} = \frac{M}{L_1} \; , \;\;\; K_{2} = \frac{\Phi_{12}}{\Phi_{22}} = \frac{M}{L_2}
\end{align}
for the general case of arbitrary transformer shapes. $K_1$ measures the proportion of the mutual magnetic flux $\Phi_{21}$ caused by a current $i_1$ in $M$ and the magnetic flux $\Phi_{11}$ caused by a current $i_1$ in $L_1$. The same principle applies to $K_2$. Note that the values of $K_\tu{i}$ can be greater than 1.
The mutual inductance $M$ is reciprocal for all linear materials with symmetric tensors for electric conductivity, permittivity and permeability~\cite{zacharias_magnetic_2022}.
Now, the definition of the coupling coefficient is given with
\begin{align} \label{eq:k}
    k = \dfrac{M}{\sqrt{L_1 L_2}} = \sqrt{K_1 K_2} \; \; . 
\end{align}

Lastly, a general expression for the quality factor $Q$ is given by the ratio of stored energy to dissipated energy per cycle: For an inductor $L$, perfect capacitors, and assuming negligible radiation losses in the \si{\kHz} region, the quality factor is
\begin{align} \label{eq:Q}
    Q = 2 \pi \frac{E_\tu{stored}}{\frac{1}{f} P_\tu{diss}} 
    = 2 \pi f \, \frac{  \displaystyle\int_0^{\hat i} L \;i \, \textrm{d}i}{\left(\frac{\hat i}{\sqrt{2}}\right)^2 R}
    = \frac{\omega L}{R} \, .
\end{align}
$R$ is the total resistance of the considered resonance~\cite{zolfaghari_stacked_2001} and $\hat i$ the total current of the inductor. 

Under the consideration of a dominant $Q$ of the coils, the three constituents of the gain in \eqref{eq:G2} are replaced for the considered \ac{ICN} and the equation yields
\begin{align} \label{eq:G1}
    G = \frac{\omega M}{R_\tu{s}} = \frac{\omega L_2}{R_\tu{s}}  \, \dfrac{M}{\sqrt{L_1 L_2}} \, \sqrt{\dfrac{L_1}{L_2}}
    = Q \, k \, n \, . 
\end{align}
Note, that the inserted $Q$ refers to the transformer with its energy stores $L_2$, $C_1$, and $C_2$, while the resonance on the right side acts like a real impedance, adding only the serial resistance $R_\tu{Tx}$ of the \ac{DFG}.

There are three conclusions due to \eqref{eq:G1}: The channel frequency should be chosen as high as reasonable, taking into account the drawback of increasing losses due to high-frequency effects in the transmit and receive chains. We consider fixed frequencies in the low \si{\kHz} range (around \SI{26}{\kHz}) and discuss the frequency choice in \autoref{sec:discussion}.
Second, a reduction of $R_\tu{s}$ is beneficial: The correct type of litz wire should be chosen~\cite{sullivan_optimal_1999}, connections within the \ac{HCR} should be kept short, and the cross-section shape of the inductor should be chosen to obtain a minimum $R$ for maximum $L$~\cite{murgatroyd_optimum_1982,murgatroyd_optimal_1989} to result in a high $Q$. Apart from minimizing $R$, the goal of maximizing $M$ requires finding a toroid with both, high $k$ and large $L$. A high turns ratio $n$ can be obtained by a dense primary winding and parallelizing the second transformer side.
Third, as the total energy dissipated must equal the energy delivered in the resonant circuit we obtain
\begin{equation}
    |P_1| \Delta t = |P_2| \Delta t ~~\Rightarrow~~ 
    \text{Re}(Z_1) = G^2  \,\text{Re}(Z_2) \; ,
\end{equation}
for a time interval $\Delta t$ at steady-state, $Z_1=v_1/i_1$ and $Z_2=v_2/i_2$. The imaginary parts is zero at resonance and the real impedance seen by the primary side equals 
\begin{equation}\label{eq:Zprim}
    \text{Re}(Z_\tu{prim})=R_1 + G^2 \, R_\tu{s} = R_1 + \frac{\omega^2 M^2}{R_\tu{s}} \; .
\end{equation}

%%%%%%%%%%%%%%%%%%%%%%%%%%%%%%%%%%%%%%%%%%%%%%%%%%%%%%%%%%%%%%%%%%%%%%%%%%%%%%%%%%%%%%%%%%%%
%%%%%%%%%%%%%%%%%%%%%%%%%%%%%%%%%%%%%%%%%%%%%%%%%%%%%%%%%%%%%%%%%%%%%%%%%%%%%%%%%%%%%%%%%%%%%
\subsection{Structural Size Limits Q}
\label{sec:theory:volume}

As argued above, and evident from \autoref{eq:G1}, we need a large secondary inductance to obtain a high transformer gain.
A large structural size is beneficial, because an increasing cross section $A$ increases the inductance faster than its reduction by the growing center radius $r_\tu{c}$, as approximated by the equation for a single-layer air-core toroid with a circular cross-section
\begin{align}\label{eq:Ltoroid}
    L_\tu{toroid} &\approx \frac{\mu_0 N^2 A}{2 \pi r_\tu{c}}  \; .
\end{align}
Here, $\mu_0$ is the vacuum permeability and dimensions are shown in \autoref{fig:Seg+THD}\,(b).  
However, a constant number of turns that is stretched around a growing cross-section will reach a point where the changes in inductance and resistance compensate. This problem and its parameter dependencies is a well known optimization problem, treated by Murgatroyd~\cite{murgatroyd_optimum_1982,murgatroyd_optimal_1989}, and an overview can be found in Ref.~\cite{murgatroyd_economic_1985}. 
$A$ can be increased until an optimum is reached where the inductance is maximum for a given total wire length $l=N \, 2\pi r$ of all turns. This is a restraint on $A$, and there exists an optimum for the ratio of outer to inner radius for a finite $l$.
The optimum number of turns for toroids with a circular cross-section and wire diameter $d$ lies at $N = 0.8165 \sqrt{l/d}$ and in turn yields the ratio $r/r_\tu{i}=(0.8165)^{-2}\approx 1.5$ and the optimal inductance
\begin{equation} \label{eq:Lopt_circ}
L_\tu{toroid,opt,circ} \approx \frac{\mu_0 \, d}{2\pi} \left( 0.2722 \left(\frac{l}{d}\right)^\frac{3}{2} + 0.25 \frac{l}{d} \right) \; ,
\end{equation}
as derived by Murgatroyd~\cite{murgatroyd_economic_1985}.

Optimal benefits are obtained from the correct area to radius ratio in combination with a large overall construction volume, coupled with a judicious selection of layers within the interior, as shown in \autoref{fig:dshape}\,(b), and explained in \autoref{sec:theory:topo:layers}.  
The upper limit is reached if the interior space is fully occupied by wires, which also affects the cross-section shape that deviates from a circle for outer layers~\cite{murgatroyd_optimal_1989,murgatroyd_optimum_2000}. Overall, increasing the dimensions of the toroid leads to an increase in $L_2$, $Q$, and thus $G$. The maximum $Q$ is defined by the available construction volume.

%%%%%%%%%%%%%%%%%%%%%%%%%%%%%%%%%%%%%%%%%%%%%%%%%%%%%%%%%%%%%%%%%%%%%%%%%%%%%%%%%%%%%%%%%%%%
%%%%%%%%%%%%%%%%%%%%%%%%%%%%%%%%%%%%%%%%%%%%%%%%%%%%%%%%%%%%%%%%%%%%%%%%%%%%%%%%%%%%%%%%%%%%
\subsection{Toroidal Transformer}
\label{sec:theory:topology}

The \ac{ICN} toroid is optimized for a finite available construction volume: optimizing the cross-sectional shape for a higher inductance, taking into account multiple layers of winding and improving coupling of the primary winding.
Then, segmentation provides a tuning mechanism that allows to change the nominal inductance with a constant inductance to resistance ratio~\cite{evans_power_1990} to a desired value.

%%%%%%%%%%%%%%%%%%%%%%%%%%%%%%%%%%%%%%%%%%%%%%%%%%%%%%%%%%%%%%%%%%%%%%%%%%%%%%%%%%%%%%%%%%%%
\subsubsection{Optimal cross-section: D-shape}
\label{sec:theory:topo:Dshape}

\begin{figure}[t!]
    \centering
    \includegraphics[width=1.0\linewidth]{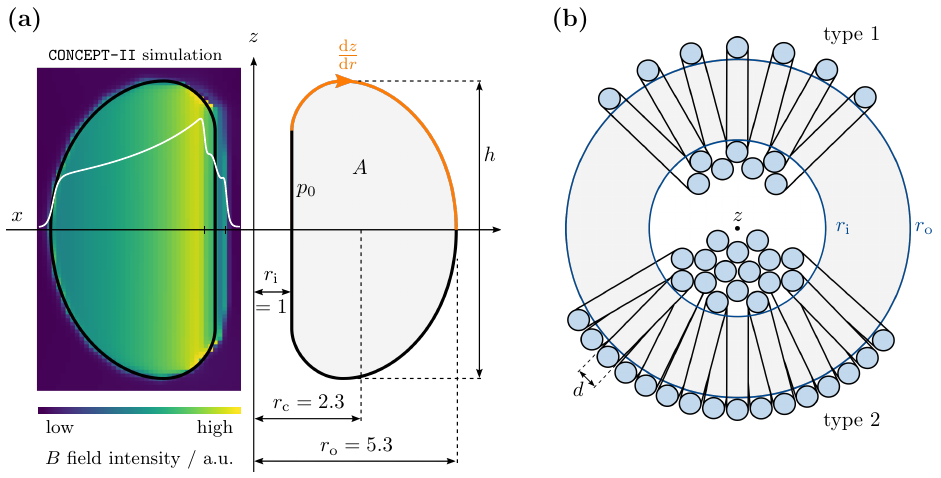}
    \caption{\label{fig:dshape}\textbf{Optimal D-shaped cross-section and multilayer toroids~\cite{murgatroyd_optimal_1989}.} In (a), the optimal D-shape cross-section is shown for the fixed optimum ratio of ${r_\tu{o}}/{r_\tu{i}}=5.3$. The left toroid-half features a field intensity plot with the field profile along the $x$-axis in white (the toroid has 3 layers on the inside, simulated in CONCEPT-II~\cite{institut_fur_theoretische_elektrotechnik_concept-ii_2023}). Highlighted in orange in the right toroid-half is the curve obtained by a stepwise evaluation and integration of \eqref{eq:step_DC}. In (b), two different types of winding arrangements of multi-layer toroidal inductors are shown.}
\end{figure}

A circular cross-section gives the largest area $A$ for a fixed turn perimeter $p_0$, but it does not provide the highest inductance for this wire length. This discrepancy is caused by the non-uniform flux density within $A$, which is denser towards the inside for toroids~\cite{shafranov_optimum_1973,murgatroyd_economic_1985}, as shown in \autoref{fig:dshape}\,(a). Due to the straight inner edge of the D-shape, it encompasses the higher magnetic flux density towards the $z$-axis of symmetry and yields approximately 15\% more inductance than the circular window for the same $p_0$~\cite{murgatroyd_economic_1985}.

For the single-layer DC loss optimized D-shape, the optimum was found to be 
\begin{align} \label{eq:LoptD}
    L_\tu{toroid,opt,D} \approx \frac{\mu_0 \, d}{2\pi} \left(0.314 \left(\frac{l}{d}\right)^\frac{3}{2} + 0.25 \frac{l}{d} \right) \;,
\end{align}
for $\frac{r_\tu{o}}{r_\tu{i}}=5.3$ and $N = 0.565 \sqrt{\frac{l}{d}}$~\cite{murgatroyd_optimal_1989}, as shown in \autoref{fig:dshape}\,(a). 
The slope (orange) is obtained by a stepwise evaluation along the radial direction, given by 
\begin{align}\label{eq:step_DC}
    \frac{\text{d} z}{\text{d} r} = \pm \dfrac{ \ln{\dfrac{\sqrt{r_\tu{i} r_\tu{o}}}{r}} }{\sqrt{ \ln{\dfrac{r}{r_\tu{i}}} \ln{\dfrac{r_\tu{o}}{r}} }} \; .
\end{align}
Integrating the slope $\text{d}z/\text{d}r$ yields coordinates for the optimized quarter section shapes.

AC losses are mainly eddy current losses due to proximity effects, since the skin effect plays only a minor role in litz wire windings. The impact by the proximity effect is twofold, internal magnetic fields due to neighboring currents cause eddy currents, but they are dominated by the second effect due to the main toroidal field through the rest of the coil, acting on the entire litz wire bundle~\cite{sullivan_optimal_1999}.
Overall, this results in eddy currents causing local non-uniform current densities that require a more complex shape optimization for high frequencies, that can be found in Ref.~\cite{murgatroyd_optimum_1982}. 
The optimum shape for high frequencies differs from a D-shape due to the stronger constraint of eddy current losses, which are highest near the center, causing the highest point of the slope to move outward. However, in the $\si{\kHz}$ range, the AC optimized shape converges to the DC optimized shape of \eqref{eq:step_DC} for litz wires with sufficiently small strands.

%%%-------------------------------------------
%%% OPTIMAL D-SHAPE only %%%
%%%-----------------------------------------

%%%%%%%%%%%%%%%%%%%%%%%%%%%%%%%%%%%%%%%%%%%%%%%%%%%%%%%%%%%%%%%%%%%%%%%%%%%%%%%%%%%%%%%%%%%%
\subsubsection{Multiple Layers}
\label{sec:theory:topo:layers}

Another way to effectively increase $L_2$, which optimizes $Q$, is by using multiple wire layers.
However, insufficient heat dissipation from inner layers becomes an issue that in turn increases copper resistance and might damage the litz wire insulation layer. To this end, we propose a single nearly dense outer layer of turns as shown in \autoref{fig:dshape}\,(b), which overlap on the inside near the axis of symmetry. This compromise gives satisfactory results and sufficient heat dissipation by air cooling. If the litz wire winding is not operated near their maximum current rating, e.g. for split-core toroids in filter stages~\cite{thieben_system_2024,mattingly_drive_2022}, multiple dense layers are an efficient option. A general study on optimal shapes for air cores and non-air core multilayered toroidal inductors can be found in Ref.~\cite{murgatroyd_optimum_2000}.

%%%%%%%%%%%%%%%%%%%%%%%%%%%%%%%%%%%%%%%%%%%%%%%%%%%%%%%%%%%%%%%%%%%%%%%%%%%%%%%%%%%%%%%%%%%%
\subsubsection{Segmentation}
\label{sec:theory:topo:segments}

The secondary side $L_2$ can be divided into $N_\tu{s}$ identical segments, which are wired in parallel but enclose a single common field, as shown in \autoref{fig:Seg+THD}\,(a) and (b). If the winding remains otherwise unchanged, both the inductance and the resistance will be reduced with $N_\tu{s}^2$, if $N_\tu{s}$ is increased.
As an example, a segmentation into two halves is considered, where one half is limited to ${N}/{N_s}={N}/{2}$ turns. Intuitively, it seems that $L$ is reduced quadratically and the resistance $R$ reduced linearly, for a dense winding that remains unchanged, due to $L\propto N^2$ via \eqref{eq:Ltoroid} and $R\propto N$ via each turn perimeter. 
However, the effect for $L$ is changed to $L\propto N$ due to the shorter magnetic core length $l$, exemplified by a long and straight solenoid with $L \propto N^2/l \propto N$ that has a constant turn density $N/l$. 
An additional slight decrease of $L$ is caused by the now open magnetic circuit, which results in less coupling of the end turns with the rest of the toroid. Mutual coupling $M$ compensates for this loss of self-inductance $L$ for toroids, since the halves now couple to the parallelized second half, i.e. to their neighbors at both ends if again arranged as a joint toroid after segmentation.
However, parallelizing equal impedances results in a $1/N_\tu{s}$ division for both $L$ and $R$. Overall, the linear contribution of the separation and the linear contribution of the parallelization result in a $1/N_\tu{s}^2$ scaling for both, $L$ and $R$.

A segmented toroid with $N_\tu{s}=12$ segments is shown in \autoref{fig:Seg+THD}\,(a) and the total current is distributed by a central node (copper plates) at the top and bottom. A lead wire of each segment is connected to the node as shown in \autoref{fig:system_overview}\,(b) for $N_\tu{s}=4$. 
Another aspect of the segmentation is that the current $i_\tu{r}$ of the \ac{HCR} is divided equally among the parallel segments, thus relaxing the copper cross section requirements for the $L_2$ winding. 
In system design situations, the construction volume is usually constrained to a maximum bounding box. For an optimal shape, $Q$ is thus fixed. Using this method, the design can start with an optimal single winding inductor and then use the segmentation to tune for a desired $L_2$ and $R_2$.

%%%%%%%%%%%%%%%%%%%%%%%%%%%%%%%%%%%%%%%%%%%%%%%%%%%%%%%%%%%%%%%%%%%%%%%%%%%%%%%%%%%%%%%%%%%%
\subsubsection{Primary Winding}
\label{sec:theory:topo:winding}

A final design decision of the toroidal topology concerns the primary winding.
For a reasonable volume and therefore limited $Q$, the primary winding should achieve a high coupling $k$, as seen in \eqref{eq:G1}. Consequently, one choice is that $L_1$ is also toroidal and (sparsely) wound around the outside of $L_2$. This ensures that the majority of the field lines are shared by both transformer sides, to maximize $k$. In \autoref{fig:Seg+THD}\,(b), such a toroidal primary winding of $L_1$ (orange) is shown for a toroid with a circular cross-section.
Known from Rogowski coils~\cite{rogowski_messung_1912} is the beneficial effect of a return wire in the $xy$-plane that counteracts the single turn of $l$ along the center circumference of the toroid to diminish the field on the outside, which is in line with the $z$-axis. For ease of fabrication, we propose a return wire along the outside at $r_\tu{o}$, which suppresses most of the stray field, but deviates from the optimal enclosed position at $r_\tu{c}$.

If a dense $L_1$ coil is used, i.e. a few turns at one point around the toroid, the result is a large $K_1$  and a very weak $K_2$. The overall $k$ is small, which is only sufficient for a design with a very large $Q$. However, such a dense $L_1$ provides several advantages for peripheral measures, like a pick-up coil to measure the current $i_\tu{r}$. Such a coil $L_3$ could be mounted on the toroid opposite to $L_1$ to focus its sensitivity locally to $L_2$ and avoid coupling between $L_1$ and $L_3$.

%%%%%%%%%%%%%%%%%%%%%%%%%%%%%%%%%%%%%%%%%%%%%%%%%%%%%%%%%%%%%%%%%%%%%%%%%%%%%%%%%%%%%%%%%%%%%
%%%%%%%%%%%%%%%%%%%%%%%%%%%%%%%%%%%%%%%%%%%%%%%%%%%%%%%%%%%%%%%%%%%%%%%%%%%%%%%%%%%%%%%%%%%%
\subsection{DFG Matching Condition}
\label{sec:theory:DFGmatch}

We have previously shown that, in order to achieve high gain for design goal~c., it is advantageous to obtain a high $Q$. Currently, the inductance $L_2$ itself was not considered. In the following, a trade-off is identified that characterizes $L_2$ in dependence of $L_\tu{Tx}$ to comply with design goal~d). This consideration arises from the specific constraints imposed by the \ac{MPI} imaging setup, in particular the circuit symmetry condition for simultaneous transmission and reception, and the absolute power consumption, which has not been considered so far.
Here, we assume a constant $Q$, which is justified for a fixed bounding box as reasoned in \autoref{sec:theory:volume}, resulting in a linear relationship of $R_2$ and $L_2$ at a fixed $\omega$. The ohmic resistances of the capacitors and the \ac{DFG} remain constant.

Our objectives are twofold, yet inherently contradictory: achieving both maximal particle signal strength at the $v_\tu{TxRx}$ port and minimal power consumption within the \ac{HCR}. The particle response is characterized by the prevalence of high-order harmonics of the resonance frequency $f_1$. These harmonics experience a significant voltage drop across an inductive voltage divider formed by the inductances $L_2$ and $L_\tu{Tx}$ when compared to the relatively minor influences of their small series resistances within the \si{\milli\ohm} range~\cite{thieben_system_2024} and the associated \ac{HCR} capacitors at harmonic frequencies.
The winding configuration of $L_\tu{Tx}$ can be conceptualized as a distributed voltage source, thereby inducing the particle voltage at the virtual ground nodes of $v_\tu{TxRx}$~\cite{sattel_setup_2015}. Ideally, in the scenario of an infinitely large $L_2$, the complete induced voltage would be dropped across $L_2$, leading to a maximal particle response at $v_\tu{TxRx}$. In contrast, a diminished ratio of $L_2/L_\tu{Tx}$ would effectively short-circuit the receive voltage, advocating a maximum value for $L_2$.
However, the cumulative series resistance $R_2 \propto N\,2\pi r$ of the toroid manifests the same increase as $L_2$ is augmented, as imposed by a constant $Q$ via \eqref{eq:Q}.
The total power consumption increases with $R_2$, but it is limited to a feasible amount. This imposes a constraint on $L_2$, because the current amplitude at resonance $i_\tu{r}$ in the \ac{HCR} must stay constant to maintain the same drive field strength.
Consequently, a partial attenuation of the particle signal is inevitable due to the voltage division effect inherent in the inductance voltage divider of $L_2$ and $L_\tu{Tx}$.

%%%-------------------------------------------
%%% DFG MATCHING CONDITION PLOTs
\begin{figure}[t!]
    \centering
    \includegraphics[width=1.0\linewidth]{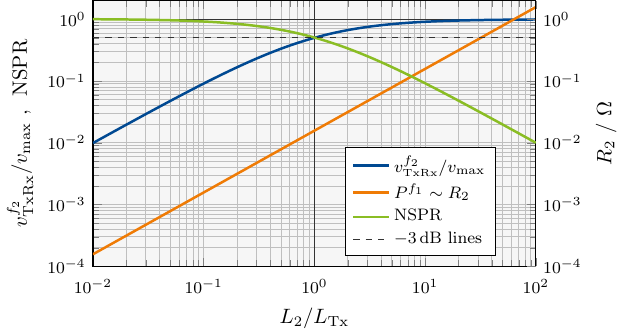}
    \caption{\label{fig:DFGmatch}\textbf{Inductance matching condition of \ac{DFG} and \ac{ICN}.} A trade-off is identified at $L_2=L_\tu{Tx}$, when the receive signal is halved (\SI{-3}{dB}), expressed by the intersection of the particle signal (blue) and the normalized signal-to-loss ratio (green). This graph demonstrates that additional power is required to avoid receive signal attenuation. }
\end{figure}
%%%-----------------------------------------

The crucial point is that the power consumption at the fundamental $f_1$ scales linearly with $R_2$ and the particle voltage at frequencies $f > f_1$ drops depending on the inductive voltage divider, when the impedance of capacitors becomes negligible. We consider a unit current $i_\tu{r}=\SI{1}{\A}$ and define the particle signal-to-power ratio (SPR) as
\begin{equation}\label{eq:SPR}
    \text{SPR} = \frac{v_\tu{TxRx}^{f_2}}{P^{f_1}} = \frac{v_\tu{TxRx}^{f_2}}{i_\tu{r}^2 \, R_2} \, .
\end{equation} 
The power consumption $P^{f_1}$ is considered at the fundamental, but the particle signal $v_\tu{TxRx}^{f_2}$ at the first harmonic. Note that for $f_2$ and all higher frequencies the inductive voltage divider is the dominant part and all harmonics are recorded.
The maximum particle voltage $v_\tu{max}$ of $v_\tu{TxRx}^{f_2}$ is defined in the limit of a large $L_2$, where it is not attenuated. To remove the influence of the absolute value of $R_2$, we normalize \eqref{eq:SPR} with its maximum in the parameter range and call it NSPR. The constituents of \eqref{eq:SPR} and the NSPR are plotted in \autoref{fig:DFGmatch}.

One possible trade-off position can be identified at \SI{-3}{\dB} of NSPR and $v_\tu{TxRx}^{f_2}$, that results in halving the receive signal (harmonics). This point coincides with $L_2=L_\tu{Tx}$, where the inductances match, which we chose as the trade-off for the design of our \ac{ICN}s. Consequently, a less attenuated particle signal requires more power.
Note that \eqref{eq:SPR} remains valid when $L_\tu{Tx}$ is varied instead of $L_2$ and the consideration of the inductance voltage divider remains identical.

%%%%%%%%%%%%%%%%%%%%%%%%%%%%%%%%%%%%%%%%%%%%%%%%%%%%%%%%%%%%%%%%%%%%%%%%%%%%%%%%%%%%%%%%%%%%
%%%%%%%%%%%%%%%%%%%%%%%%%%%%%%%%%%%%%%%%%%%%%%%%%%%%%%%%%%%%%%%%%%%%%%%%%%%%%%%%%%%%%%%%%%%%
\subsection{Channel Decoupling}
\label{sec:theory:decoupling}

The use of multiple channels in \ac{MPI} has the advantage of simultaneous sampling of a 3D \ac{FOV}.
Due to imperfect spatial orthogonality of these channels and close proximity of resonance frequencies, typically within a range of less than $\Delta f = 0.03 f_1$~\cite{knopp_magnetic_2017}, strong coupling between the different drive-field channels is expected. Thus even weak coupling coefficients presents a particular challenge, especially in case of resonant circuits with high $Q$, necessitating the development of effective decoupling mechanisms with introducing minimal additional resistance to the \ac{HCR}.

Causes of channel coupling are misalignment of the drive-field coils, non-orthogonal field components, and undesired loops in the connecting wires. Small errors in positioning, e.g. for two planar coils with $k\propto \cos(\beta)$, can lead to coupling with significant currents. A $\beta=\SI{3}{\degree}$ results in $k=0.052$, which can generate a current $i_\tu{r,2}$ in channel 2 of similar magnitude to the original current $i_\tu{r,1}$ in channel 1. 
%, for reasonable values of inductance and $Q$. 
For this purpose, if we consider the complex impedance $Z_\tu{HCR,2}$ of the second channel's \ac{HCR} and let $j$ be the imaginary unit, then the current $i_\tu{r,2}$ in the second channel at the first angular frequency $\omega_\tu{r,1}$ is caused by induction via $i_\tu{r,1}$ and $M$, as in
\begin{align}\nonumber
    \frac{i_\tu{r,2}(\omega_\tu{r,1})}{i_\tu{r,1}(\omega_\tu{r,1})} 
    &= \frac{v_\tu{ind,21}(\omega_\tu{r,1})}{Z_\tu{HCR,2}(\omega_\tu{r,1})}\frac{1}{i_\tu{r,1}(\omega_\tu{r,1})} \\\label{eq:decoup}
    &= \frac{j \omega_\tu{r,1} M \, i_\tu{r,1}(\omega_\tu{r,1})}{Z_\tu{HCR,2}(\Delta\omega+\omega_\tu{r,2}) \, i_\tu{r,1}(\omega_\tu{r,1})} \\\nonumber
    &\approx \frac{j \omega_\tu{r,1} M}{j 2 \Delta\omega (L_\tu{Tx,2}+L_\tu{2,2})}
    \approx \frac{\omega_\tu{r,1} M}{4 \Delta\omega L_\tu{2,2}} = \frac{1}{4} \frac{\omega_\tu{r,1}}{\Delta\omega} \, k \, .
\end{align}
Note that the matching criterion of \autoref{sec:theory:DFGmatch} is respected with $L_\tu{Tx,2} = L_\tu{2,2}$ and we approximate $Z_\tu{HCR,2}$ near $\omega_\tu{r,2}$ with $j2\Delta\omega L_{2,2}$ as shown in the appendix \ref{sec:app:wL}.
%dominant imaginary part and low Rs2 near w2!
With $\tfrac{\Delta\omega}{\omega_\tu{r,1}} = \tfrac{(\omega_\tu{r,1}-\omega_\tu{r,2})}{\omega_\tu{r,1}} = \tfrac{1}{75}$ (see beginning of \autoref{sec:results}) and the aforementioned misalignment of $\beta=\SI{3}{\degree}$, this example results in $\tfrac{i_\tu{r,2}}{i_\tu{r,1}} = 0.25 \cdot 75 \cdot 0.052 = 0.975 \approx 1$.
The consequence is a severe distortion of the Lissajous trajectory, which is already significant for values ${i_\tu{r,2}}/{i_\tu{r,1}} \geq \SI{10}{\%}$~\cite{von_gladiss_influence_2018}.

In order to avoid the negative effects of uncompensated coupling, such as trajectory distortion, additional power dissipation, detuning, and frequency beating that will act on the power amplifier, we consider 3 types of decoupling schemes: capacitive, inductive and active compensation. 
Capacitive compensation is narrowband and requires additional connections between channels, but the \ac{ESR} of capacitors is low. 
Inductive compensation is directed at counteracting $M$ and is broad-band, but additional coils typically increase the series resistance, thus reducing $Q$.
Finally, active decoupling uses the channel's amplifiers, but demands reactive power in the band-pass filters (\autoref{fig:system_overview}\,(a)) and for the miss-matched \ac{HCR}. Further, it requires accurate feedback for current control. It should be considered a last resort, as the amplifier will see a reactive load at other frequencies, and maximum voltage ratings may be exceeded over the course of a full Lissajous trajectory cycle.

\begin{figure}[t!]
    \centering
    \includegraphics[width=1.0\linewidth]{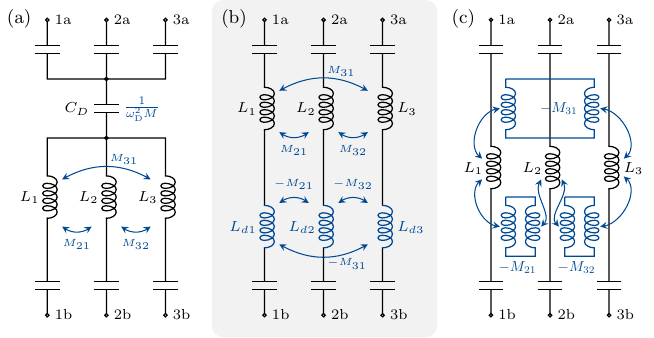}
    \caption{\label{fig:decoupling_schematic}\textbf{Schematic of decoupling schemes with 3 TxRx channels.} The schematics refer to one half of the symmetric \ac{HCR}, where $v_\tu{TxRx}$ can be tapped across each numbered a-b terminal pair (compare \autoref{fig:system_overview}). Capacitive decoupling for a single frequency via a common capacitor $C_\tu{D}$ is shown in (a), with the condition of $M_{21}\approx M_{32} \approx M_{31}$. Inductive decoupling strategies (broadband) are shown in (b) and (c), for the case of series inductors and the case of separate windings, respectively. }
\end{figure}

In general, capacitive decoupling may not offer an exact solution, and there are unsolvable combinations of coupling depending on the sign of the coupling coefficients. However, capacitive compensation is advantageous for the particle signal of a transmit-receive circuit because a large common series capacitor can be used, which becomes a short at harmonic frequencies and has a very low series resistance compared to coil windings. This requires two common nodes between channels and similar inductive coupling coefficients of the same sign, due to the single narrowband decoupling point. 
We could use 3 capacitors, 2 within each channel or alternatively one common capacitor for all channels.
The mentioned low differences of drive-field frequencies in \ac{MPI} are a premise, and a limitation is that the sum of all currents flows through this capacitor. 
A schematic is shown in \autoref{fig:decoupling_schematic}\;(a), which represents only the right part of the symmetric \ac{HCR}, with $v_\tu{TxRx}$ across each numbered a-b terminal pair, shown for 3 channels each.

%%%-------------------------------------------
%%% DECOUPLING schematic %%%
%%%-----------------------------------------

Inductive compensation is a general alternative for all coupling coefficient signs: Either a coil is added in series to $L_\tu{Tx}$ for each unique off-diagonal entry of the system's impedance matrix as in \autoref{fig:decoupling_schematic}\,(b), or the same number of separate windings is used to introduce the desired compensation without a galvanic connection by careful positioning as shown in \autoref{fig:decoupling_schematic}\,(c). A disadvantage of the first technique is the additional high current lines, stray fields, their influence on the tuning of the resonators, additional series resistance, as well as their mounting and cooling effort. 
The second technique uses separate windings that are sensitive to a partial amount of the drive-field, equal to the $M$ introduced by coupling. A suitable inductive decoupling position is the \ac{ICN} itself, as implied in the last paragraph of \autoref{sec:theory:topo:winding}. Turns on the outer surface of the toroid are only capable of utilizing integer multiples to achieve a match with $M$. However, a slab outfitted with a wire loop can be introduced to encompass a partial quantity of the magnetic field inside: a loop on a slab that is inserted into a prepared gap within the toroid. Adjustment of the slabs can be used to fine-tune each channel if the \ac{ICN} was prepared with such gaps. The sign can be adapted by inverting the orientation of the wire-loop. However, the principle remains the same if it is done with partial fields in the proximity of the \ac{DFG}.

For our two-channel human-sized system~\cite{thieben_system_2024}, we decided to use a single common capacitor that carries a peak current of $i_\tu{r,1}+i_\tu{r,2}$ (depending on the phase of the Lissajous cycle), introducing the same voltage as $M$ into each circuit, but with opposite sign.
The common decoupling capacitor $C_\tu{D}$ is calculated at a single frequency $\omega_\tu{D}$ in-between both drive-field frequencies by
\begin{equation}
    C_\tu{D} = \frac{1}{\omega_\tu{D}^2 M} = \frac{1}{(\pi (f_\tu{r,2}+ f_\tu{r,1}))^2\; k \sqrt{L_\tu{Tx,1} L_\tu{Tx,2}}}
\end{equation}
This narrowband solution is acceptable for our implemented two-channel system, and the additional \ac{ESR} introduced by the decoupling capacitor is small.

%%%%%%%%%%%%%%%%%%%%%%%%%%%%%%%%%%%%%%%%%%%%%%%%%%%%%%%%%%%%%%%%%%%%%%%%%%%%%%%%%%%%%%%%%%%%
%%%%%%%%%%%%%%%%%%%%%%%%%%%%%%%%%%%%%%%%%%%%%%%%%%%%%%%%%%%%%%%%%%%%%%%%%%%%%%%%%%%%%%%%%%%%
%%%%%%%%%%%%%%%%%%%%%%%%%%%%%%%%%%%%%%%%%%%%%%%%%%%%%%%%%%%%%%%%%%%%%%%%%%%%%%%%%%%%%%%%%%%%
\section{Methods and Implementation}
\label{sec:methods}

Guided by the theory to optimize the \ac{ICN} for a given bounding box, we now describe the implementation and name the changes to our design that deviate from the stated optimum.
We employ simulations to assess performance in \autoref{sec:methods:simu}, which include a D-shaped toroid and are used to refine our decision on the number of segments and the type of primary winding. 
The construction of two \ac{ICN}s is described in \autoref{sec:methods:construction} and the measurement methodology to analyse the prototypes is given in \autoref{sec:methods:meas}.

%%%%%%%%%%%%%%%%%%%%%%%%%%%%%%%%%%%%%%%%%%%%%%%%%%%%%%%%%%%%%%%%%%%%%%%%%%%%%%%%%%%%%%%%%%%%
%%%%%%%%%%%%%%%%%%%%%%%%%%%%%%%%%%%%%%%%%%%%%%%%%%%%%%%%%%%%%%%%%%%%%%%%%%%%%%%%%%%%%%%%%%%%
\subsection{Simulations}
\label{sec:methods:simu}

Linear circuit simulations are performed with LTspice 17.1 (Analog Devices, MA, USA)~\cite{analog_devices_inc_ltspice_2023}. Simulations of the magnetic field and the transformer's impedance matrix are performed using CONCEPT-II (Institut f\"ur Theoretische Elektrotechnik, Hamburg University of Technology, Germany)~\cite{institut_fur_theoretische_elektrotechnik_concept-ii_2023}.

\subsubsection{Toroid Geometry}
The CONCEPT-II software is based on the \ac{MoM} that solves electromagnetic boundary or volume integral equations in the frequency domain~\cite{gibson_method_2008}. It is especially suited for the numerical computation of 3D radiation and scattering problems. 
CONCEPT-II is used in our work to calculate and test different toroidal transformer configurations.
Input parameters include the general geometry (distances, shape and size of $A$, number of segments $N_\tu{s}$, turns $N_1$ and $N_2$, primary winding shape, number of overlapping layers inside the toroid), while important output parameters include the impedance matrix $\mv{Z}$, 3D field plots and conductors current densities.
A two-port network is used to estimate $k$ and $G$ via the impedance matrix of the transformer
\begin{equation}\label{eq:Zmat}
    \mv{Z} =  
    \begin{bmatrix}
    Z_{11} & Z_{12} \\
    Z_{21} & Z_{22}
    \end{bmatrix}
    \approx \begin{bmatrix}
    R_\tu{1} + j \omega L_1 & j \omega M \\
    j \omega M  & R_\tu{2} + j \omega L_2
    \end{bmatrix} \, . 
\end{equation}
Different numbers of segments $N_\tu{s}$ are simulated as well as the coupling between neighboring segments. Stray fields are minimized and coupling is maximized by looking at different winding options for $L_1$ (dense vs. sparse/distributed), the suited number of turns $N_1$, and the current density in the copper-plates where parallel segments are joined. 
Litz wire windings are approximated by a single thin wire, assuming that the currents are restricted to the direction of the wire, and copper conductivity of surfaces and wires is set to $\sigma = \SI{58e6}{\siemens\per\m}$.
The results of the simulation are used to adapt the design and finally to compare the expectations with the manufactured prototypes in terms of $L$, $k$, $n$, and $G$.
One simulated $B$-field plot is shown in \autoref{fig:dshape}\,(a) for a number of 3 overlapping wire layers on the inside. One complete model (4 segments) is shown within the simulation framework in \autoref{fig:system_overview}\,(b), including copper distribution plates.

\subsubsection{Circuit Analysis}
With simulated values for $L_1$, $L_2$, $k$ and resistors as input parameters, the resonance behavior is analyzed in the schematic circuit analyzer tool LTspice. A model of the entire \ac{HCR} is simulated, including the band-pass filter stage, the \ac{ICN} and two channels including crosstalk.
Different decoupling strategies are tested to probe the schematic, based on \autoref{fig:ECD}, and to tune values for decoupling elements. The input impedance value $Z_\tu{prim}$ was also simulated with LTspice. A simplified LTspice model without the band-pass is provided in the supplements with two coupled channels including the \ac{HCR} and approximate values for capacitors and inductors.

%%%%%%%%%%%%%%%%%%%%%%%%%%%%%%%%%%%%%%%%%%%%%%%%%%%%%%%%%%%%%%%%%%%%%%%%%%%%%%%%%%%%%%%%%%%%
%##################################################################################
\begin{table*}[t!]
\centering
\caption{\label{tab:results}Values from simulation (yellow) and measurement (blue) for both constructed ICNs. Bold font denotes important values.}
\resizebox{1.0\textwidth}{!}{
\begin{tblr}{
    colspec = {ccc|c|c|r|c|c|c|c|c|c|c|c||c|c},
    row{2} = {ukesec2!10},
    row{3} = {ukesec2!10},
    row{4} = {ibilight!20},
    row{5} = {ibilight!20}, 
    row{6} = {ukesec2!10},
    row{7} = {ukesec2!10},
    row{8} = {ibilight!20},
    row{9} = {ibilight!20}, 
    cell{1}{1} = {white},
    cell{2}{1} = {white},
    cell{6}{1} = {white},
    cell{7}{1} = {white},
    column{15} = {white},
    column{16} = {white},
  }
& & & $L$\,/\,\si{\micro\henry} 
& $R$\,/\,\si{\milli\ohm}  & $N$~~ & $f_1$\,/\,\si{Hz}  & $M$\,/\,\si{\micro\henry} & $Q_\tu{ICN} $ & $Q_\tu{HCR}$ & $n$ & $k$ & $G$ & $Z_\tu{prim}$\,/\,\si{\ohm} & $L_\tu{Tx}$\,/\,\si{\micro\henry} & $R_\tu{Tx}$\,/\,\si{\milli\ohm} 
%%-------------------------------
%% ICN 1
\\\hline
\SetCell[r=4]{c} \rotatebox[origin=c]{90}{\text{\normalsize ICN~1}}
& \SetCell[r=2]{c} \rotatebox[origin=c]{90}{\text{sim.}} & pri.
& 6.21 & 200 & 12
& \SetCell[r=2]{c} 25699   
& \SetCell[r=2]{c} 5.47   
& \SetCell[r=2]{c} 219   
& \SetCell[r=2]{c} 90.1  
& \SetCell[r=2]{c} 0.62  
& \SetCell[r=2]{c} 0.54   
& \SetCell[r=2]{c} 30.2
& \SetCell[r=2]{c} 26.8 &  &
\\
& & sec. & \textbf{16.3} & 12.0 & $4\cdot 36$ & & & & & & & & & &                 
\\[4pt]
& \SetCell[r=2]{c} \rotatebox[origin=c]{90}{\text{meas.\vphantom{i}}} & pri.
& 6.65 & 233 & 12  
& \SetCell[r=2]{c} 25699
& \SetCell[r=2]{c} 6.08  
& \SetCell[r=2]{c} \textbf{215} 
& \SetCell[r=2]{c} 90.8
& \SetCell[r=2]{c} 0.63 
& \SetCell[r=2]{c} 0.57  
& \SetCell[r=2]{c} \textbf{33.0} 
& \SetCell[r=2]{c} 32.6 
& \SetCell[r=2]{c} 14.4  & \SetCell[r=2]{c} 17.2  
\\
& & sec. & \textbf{16.7} & 12.5 & $4 \cdot 36$ & & & & & & & & & &  
\\\hline
%%-------------------------------
%% ICN 2
\SetCell[r=4]{c} \rotatebox[origin=c]{90}{\text{\normalsize ICN~2}} 
& \SetCell[r=2]{c} \rotatebox[origin=c]{90}{\text{sim.}} & pri.
& 6.88  & 200  & 13
& \SetCell[r=2]{c} 26042
& \SetCell[r=2]{c} 4.09 
& \SetCell[r=2]{c} 230
& \SetCell[r=2]{c} 77.1 
& \SetCell[r=2]{c} 0.94 
& \SetCell[r=2]{c} 0.56 
& \SetCell[r=2]{c} 40.8
& \SetCell[r=2]{c} 27.6  &  &
\\
& & sec. & \textbf{7.73} & 5.5 & $6\cdot 24$ & & & & & & & & & &                
\\[4pt]
& \SetCell[r=2]{c} \rotatebox[origin=c]{90}{\text{meas.\vphantom{i}}} & pri.
& 6.85 & 210 & 13  
& \SetCell[r=2]{c} 26042
& \SetCell[r=2]{c} 4.39 
& \SetCell[r=2]{c} \textbf{220}
& \SetCell[r=2]{c} 76.4
& \SetCell[r=2]{c} 0.94 
& \SetCell[r=2]{c} 0.60 
& \SetCell[r=2]{c} \textbf{43.0}
& \SetCell[r=2]{c} 31.1 
& \SetCell[r=2]{c} 9.74  & \SetCell[r=2]{c} 10.9 
\\
& & sec. & \textbf{7.8} & 5.8 & $6\cdot 24$ & & & & & & & & & &
\end{tblr}
}%resizebox
\end{table*}
%##################################################################################
%%%%%%%%%%%%%%%%%%%%%%%%%%%%%%%%%%%%%%%%%%%%%%%%%%%%%%%%%%%%%%%%%%%%%%%%%%%%%%%%%%%%%%%%%%%%

%%%-------------------------------------------
%%% ICN result images %%%
\begin{figure*}[ht]
    \centering
    \includegraphics[width=0.85\linewidth]{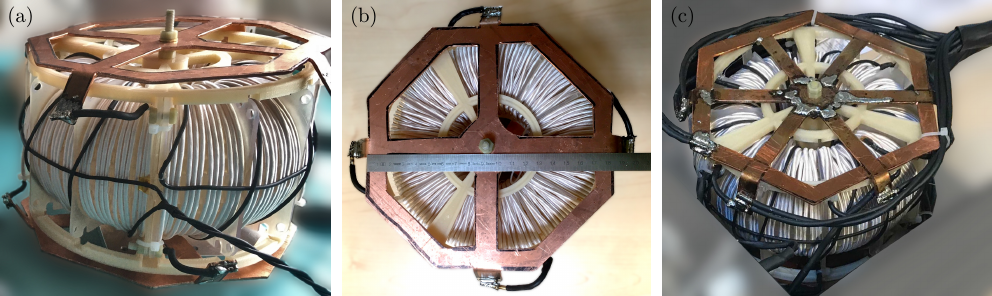}
    \caption{\label{fig:resultsICN}\textbf{Two ICNs for a two channel MPI system.} In (a), the ICN~1 with $N_\tu{s}=4$ segments is shown (primary winding in black) and in (b) the top of the same ICN is shown with a reference scale in cm (without primary winding). ICN~2 with $N_\tu{s}=6$ segments is shown in (c), after installation. The connecting litz wire of the \ac{HCR} and the primary winding are insulated with black shrinking tube.
    }
\end{figure*}
%%%-----------------------------------------

%%%%%%%%%%%%%%%%%%%%%%%%%%%%%%%%%%%%%%%%%%%%%%%%%%%%%%%%%%%%%%%%%%%%%%%%%%%%%%%%%%%%%%%%%%%%
%%%%%%%%%%%%%%%%%%%%%%%%%%%%%%%%%%%%%%%%%%%%%%%%%%%%%%%%%%%%%%%%%%%%%%%%%%%%%%%%%%%%%%%%%%%%
\subsection{Construction of two Toroidal ICNs}
\label{sec:methods:construction}

To weigh different design decisions of the \ac{ICN}, we followed the reasoning of \autoref{sec:theory} and \ref{sec:methods:simu} consecutively. The construction was performed after the simulation, with the probed values for parameters $L$, $N$, the number of segments $N_\tu{s,1}=4$ and $N_\tu{s,2}=6$ as given in \autoref{tab:results}, and the chosen dimensions of $r_\tu{i}=\SI{4}{\cm}$, $r_\tu{o}=\SI{9.2}{\cm}$, and the height $h=\SI{8.6}{\cm}$ for the D-shaped second side. Moreover, we use the channel frequencies $f_{1,1} = \SI{25.699}{\kHz}$ and $f_{1,2} = \SI{26.042}{\kHz}$ with a distance of $\Delta f=\SI{343}{\Hz}$. 

The choice of $N_\tu{s}=6$ segments for the second channel is owed to the fact that we implemented a saddle coil (\ac{DFG} of channel 2), which requires a higher current than the first channel to meet field specifications~\cite{thieben_system_2024}, thus we choose to reduce $R_2$. Due to the fact that $Q$ remains constant, the increase in gain is caused by the turns ratio $n$ due to the shift in the ratio of $L_1$ to $L_2$.
As reasoned in \autoref{sec:theory:topo:segments}, the inductance decreases for a higher number of segments, which also better fits to the lower \ac{DFG} inductance of the saddle coil to comply with the matching condition of \autoref{sec:theory:DFGmatch}. 
To further increase the turns ratio $n$ of the second channel, we chose $N_{1,2}=13$ instead of $N_{1,1}=12$.
As a consequence, we were able to use the same support structure for winding and both \ac{ICN}s are of equal dimensions in spite of their different objectives.
The 3D printed toroidal support structure of the \ac{ICN} is made from the high temperature resin RS-F2-HTAM-02 (Formlabs, MA, USA). Air cooling along the symmetry axis is installed using shielded fans, to prevent impedance changes by insufficient heat dissipation.

A difference between the derived optimal design and the constructed \ac{ICN}s is the ratio of radii $r_\tu{i}/r_\tu{o} = 2.3$ for the D-shaped toroid. This deviation to the optimum of $r_\tu{i}/r_\tu{o} = 5.3$ is a compromise to gain more space for a feasible winding through the center of the toroid, to reduce the amount of inner layers, to increase the air cooling surface, and to comply with a limit on the available construction height. Basically, we calculated the optimal design to fully utilize the available construction space height $h$, and then opted to increase $r_\tu{i}$ while keeping the other parameters constant.

Regarding the primary winding, a dense (localized) winding is unsuited due to a low $k$, as reasoned in \autoref{sec:theory:topo:winding}. Sparse windings on a helical path yielded best results, thereby forming a toroid on top on the outer surface around the $L_2$ toroid, as shown in \autoref{fig:Seg+THD}\,(a), (b) and \autoref{fig:resultsICN}\,(a).

To minimize AC losses within the conducting material, a silk-wrapped litz wire of $2000$ strands of \SI{50}{\micro\meter} copper (effective copper cross-section of \SI{3.1}{\mm^2}) is used for the secondary transformer winding. The primary winding consists of $400$ strands of \SI{50}{\micro\meter} copper (effective copper cross-section of \SI{0.78}{\mm^2}) and is wrapped in shrinking tube to increase durability and breakdown voltage.
Parallel segments are connected on a \SI{1}{\mm} thick copper plate (top and bottom), which serves primarily as a distribution platform and to connect parallel litz wires of the \ac{HCR} and the \ac{ICN} with a solder connection. Also, balancing currents can be equalized on this low resistance copper plate.

%%%%%%%%%%%%%%%%%%%%%%%%%%%%%%%%%%%%%%%%%%%%%%%%%%%%%%%%%%%%%%%%%%%%%%%%%%%%%%%%%%%%%%%%%%%%
%%%%%%%%%%%%%%%%%%%%%%%%%%%%%%%%%%%%%%%%%%%%%%%%%%%%%%%%%%%%%%%%%%%%%%%%%%%%%%%%%%%%%%%%%%%%
\subsection{Measurements}
\label{sec:methods:meas}

Measurements of inductors are performed with the LCR meter \textit{Keysight E4980AL} (Keysight Technologies, CA, USA) at the channel's frequency. The coupling coefficient of a transformer is measured by using the short and open circuit inductances~\cite{mit_department_of_electrical_engineering_magnetic_1977} in
\begin{equation}\label{eq:kmeas}
    k \approx \sqrt{1-\frac{ \left. L_\textup{1} \right|_{L_2\,\text{short}} }{ \left. L_\textup{1} \right|_{L_2\,\text{open}}}} \; .
\end{equation}
Note, that $L_2$ needs to be sufficiently shorted (more difficult at high frequencies) and the quality factor of the secondary side should be $Q>50$ for this model to be accurate. A comparison of different measurement techniques at higher leakage can be found in Ref.~\cite{hayes_inductance_2003}.

The LCR meter is used for \autoref{tab:results} to determine $L$, the series resistance $R$, and $k$ in the measured rows. All other values in these rows are calculated: $M$ via \eqref{eq:k}, $Q_\tu{ICN}$ via \eqref{eq:Q} with $L$ and $R_2$ of the same row, likewise $Q_\tu{HCR}$ with $L$ and $R_\tu{s}=R_2+R_\tu{Tx}$, $n$ via \eqref{eq:n} using the primary and secondary measured $L$, and $G$ via \eqref{eq:G1} for each row. For the simulated rows of \autoref{tab:results}, $L$ and $M$ were obtained from the simulation in CONCEPT-II, and $k$ was calculated via \eqref{eq:k} here .

%%%%%%%%%%%%%%%%%%%%%%%%%%%%%%%%%%%%%%%%%%%%%%%%%%%%%%%%%%%%%%%%%%%%%%%%%%%%%%%%%%%%%%%%%%%%
%%%%%%%%%%%%%%%%%%%%%%%%%%%%%%%%%%%%%%%%%%%%%%%%%%%%%%%%%%%%%%%%%%%%%%%%%%%%%%%%%%%%%%%%%%%%
%%%%%%%%%%%%%%%%%%%%%%%%%%%%%%%%%%%%%%%%%%%%%%%%%%%%%%%%%%%%%%%%%%%%%%%%%%%%%%%%%%%%%%%%%%%%
\section{Results}
\label{sec:results}

We have designed, simulated and fabricated two \ac{ICN}s with $f_{r_1} = \SI{25.699}{\kHz}$ ($N_\tu{s}=4$) and $f_{r_2} = \SI{26.042}{\kHz}$ ($N_\tu{s}=6$) for a two channel human-sized \ac{MPI} system and integrated both into our MPI scanner~\cite{thieben_system_2024}. The simulation and measurement results of the final design are summarized in \autoref{tab:results} and details of the construction are described in \autoref{sec:methods:construction}. Pictures of both constructed \ac{ICN}s are shown in \autoref{fig:resultsICN}. 

The deviation between measured and simulated inductances $L_2$ is below \SI{2.5}{\%} and below \SI{1}{\%} for the first and second \ac{ICN}, respectively.
The measured value for $k$ is about \SIrange{5}{7}{\%} larger than the simulated value. Therefore $M$ is also larger in rows 2 and 4, because its calculation is based on $k$ and $L_2$. This results in the measured $G$ being \SI{9.3}{\%} and \SI{5.4}{\%} larger for \ac{ICN} 1 and 2, respectively. $Q_\tu{ICN}$ refers only to the self-inductance $L_2$ and series $R_2$ of the secondary transformer side, while $Q_\tu{HCR}$ is the more important measure that includes the entire resonant load. Here, $R_\tu{s}=R_2+R_\tu{Tx}$ represent the losses of the \ac{HCR} and resemble the gain actually achieved. Later current measurements showed a gain very similar to these values, although they are marginally lower due to losses of additional connections and wires (e.g. decoupling capacitor). 

The inductors of \ac{DFG} and \ac{ICN} are nearly matched for both channels as explained in \autoref{sec:theory:DFGmatch}, with $L_\tu{Tx,1} = \SI{14.4}{\micro\henry} \approx L_{2,1} = \SI{16.7}{\micro\henry}$ for channel 1, and with $L_\tu{Tx,2} = \SI{9.74}{\micro\henry} \approx L_{2,2} = \SI{7.8}{\micro\henry}$ for channel 2. 
Note, that the target input impedance $Z_\tu{prim}$ of \eqref{eq:Zprim} remains for both \ac{ICN}s at around \SI{30}{\ohm}, which is the load after the band-pass filter. Also, $Q$ of both \ac{ICN}s remains constant with $215\approx220$ (measured values) for the different segmentation, as argued in \autoref{sec:theory:topology}. The increase in $G$ is achieved by augmentation of $n$ due to the segmentation.

Due to the non-orthogonal \ac{DFG} channels, there is a residual coupling of $k=0.062$~\cite{thieben_system_2024}, which has serious implications for both resonators, as described in \autoref{sec:theory:decoupling}. The channels were decoupled by \SI{-35}{\dB} using a single common decoupling capacitor carrying both currents $i_\tu{r,1}$ and $i_\tu{r,2}$. This choice provided good results with minimal changes to the existing \ac{HCR}, without introducing a large series resistance and with high temperature stability.

%%%%%%%%%%%%%%%%%%%%%%%%%%%%%%%%%%%%%%%%%%%%%%%%%%%%%%%%%%%%%%%%%%%%%%%%%%%%%%%%%%%%%%%%%%%%
%%%%%%%%%%%%%%%%%%%%%%%%%%%%%%%%%%%%%%%%%%%%%%%%%%%%%%%%%%%%%%%%%%%%%%%%%%%%%%%%%%%%%%%%%%%%
\section{Discussion}
\label{sec:discussion}

%%%%% General 
Applications that require highly linear transformers, such as \ac{MPI}, rely on circuits that do not utilize magnetic materials that saturate. Any harmonic distortion may obscure the weak receive signal and degrade image quality. As a result, air-core structures are chosen at the expense of increasing component dimensions, which sets the maximum $Q$ for an available construction volume.
%%%%% Inhaltlich
In this study, we present a blueprint for a high-gain linear transformer with a high quality factor. We formulate an expression for the gain as a function of mutual inductance, optimize the cross-sectional shape, employ segmentation to shift the nominal inductance, deduce a primary winding topology, and incorporate multiple layers to enhance performance. In the context of the emerging imaging modality \ac{MPI}, we establish a matching condition that balances particle signal and power consumption. Further, we elaborate on various decoupling techniques for multichannel systems and support our assessment of the \ac{ICN} with simulations.

In terms of human safety, the entire \ac{HCR} has floating potentials and component voltages in patient proximity are reduced by the \ac{ICN} due to its high current gain that drives the low-inductance \ac{DFG} to generate the required \ac{MPI} drive-field.
The symmetric design of the \ac{HCR} allows for fundamental filtering at the signal tap, but results in partial receive signal attenuation due to the inductive voltage divider of $L_2$ and $L_\tu{Tx}$.
The TxRx topology generally reduces power consumption by saving space through the elimination of dedicated receive coils, but consumption is approximately doubled by the \ac{ICN}. In addition, common decoupling strategies such as gradiometric receive coils reduce the design requirements for linearity within the transmit chain. A direct comparison in signal quality between this TxRx topology and a gradiometric receive topology is currently pending, although other TxRx systems have been characterized~\cite{paysen_characterization_2020}.

%%%%% Rahmenbedingungen
The presented toroidal transformer blueprint is limited by its restriction to the quasi-static regime of electromagnetics, where wave propagation effects are not dominant. Additionally, if AC losses due to proximity effects dominate in the toroid, the D-shape optimization will result in a different shape~\cite{murgatroyd_optimum_1982}. 
Parameters given or assumed in this study, such as frequency or volume, require careful selection: $Q$ increases linearly with frequency and should be chosen as high as possible, taking into account component voltages, high reactive power (in resonance)~\cite{schmale_design_2015}, signal induction, wave propagation effects at high harmonics, required receive bandwidth, \ac{PNS} and \ac{SAR} limits~\cite{bohnert_effects_2010,schmale_mpi_2015,schmale_human_2013,saritas_magnetostimulation_2013,grau-ruiz_magneto-stimulation_2022}.
%%%%% DECISION for the design
Our decision regarding the deviation of ${r_\tu{o}}/{r_\tu{i}}$ from the optimum was aimed at facilitating the winding process and reducing the number of stacked inner layers.
This decision allowed winding a nearly dense outer layer with 2 to 3 stacked inner layers. Regarding \eqref{eq:Ltoroid} with a given constant area and a linearly increasing $r_\tu{c}$, we assume that the deviation from the optimal inductance is small. The copper plates at the top and bottom (nodes of parallel segments) should be designed to facilitate air cooling of the inner layers.

%%%%% KEY Insights gained/REVEALED by our analysis
A key insight is that the available construction space should be exhausted and the quality factor of the \ac{ICN} benefits from a large volume, yielding a high gain. 
Therefore, the construction volume should be as large as cost, weight, and size factors will allow. If a lower $L_2$ is desired compared to a single winding on the maximized toroid, segmentation provides the means to reduce both inductance and resistance at the same rate with a constant $Q$. Note, this also increases the turns ratio $n$, causing a higher gain $G$. Independent of geometric choices, the method of moment based simulation provided accurate results and the simulated inductance value deviated less than \SI{2.5}{\percent} from the manufactured \ac{ICN}s. 
Two linear \ac{ICN}s were built based on our presented schematic optimization that feed a floating 2-channel \ac{HCR}, including crosstalk decoupling, for a human-sized \ac{MPI} head scanner~\cite{thieben_system_2024} to fulfill safety precautions on a path towards the clinical integration of \ac{MPI}.

%%%%%%%%%%%%%%%%%%%%%%%%%%%%%%%%%%%%%%%%%%%%%%%%%%%%%%%%%%%%%%%%%%%%%%%%%%%%%%%%%%%%%%%%%%%%
%%%%%%%%%%%%%%%%%%%%%%%%%%%%%%%%%%%%%%%%%%%%%%%%%%%%%%%%%%%%%%%%%%%%%%%%%%%%%%%%%%%%%%%%%%%%
\begin{acknowledgments}
We are very grateful to Christian Findeklee for discussions, proofreading, and initial input on variations of resonant circuit decoupling.
The authors would also like to thank Heinz-D. Br\"uns and Christian Schuster of the Institut f\"ur Theoretische Elektrotechnik from the Hamburg University of Technology for providing the CONCEPT-II software and assisting with questions regarding simulations.\\
This work was supported by the German Research Foundation (DFG, grant numbers GR 5287/2-1, KN 1108/7-1), the Forschungszentrum Medizintechnik Hamburg (grant number 01fmthh2018). The Fraunhofer IMTE is supported by the EU (EFRE) and the State Schleswig-Holstein, Germany (Project: IMTE – Grant: 124 20 002 / LPW-E1.1.1/1536).
\end{acknowledgments}

%%%%%%%%%%%%%%%%%%%%%%%%%%%%%%%%%%%%%%%%%%%%%%%%%%%%%%%%%%%%%%%%%%%%%%%%%%%%%%%%%%%%%%%%%%%%
%%%%%%%%%%%%%%%%%%%%%%%%%%%%%%%%%%%%%%%%%%%%%%%%%%%%%%%%%%%%%%%%%%%%%%%%%%%%%%%%%%%%%%%%%%%%
\section*{Author Contributions}

F.M. F.F., F.T., T.K., and M.G. contributed to the conceptualization and theory.
F.M. and F.F. performed the simulations and measurements.
F.M., F.F., F.T., and M.G. constructed the MPI components. 
I.S. contributed to the multi-channel circuit analysis and decoupling theory.
T.K. and M.G. supervised the project. 
F.M. wrote the original draft with support of M.M., M.G., F.F., F.T., I.S. and T.K. All authors reviewed the manuscript.

%%%%%%%%%%%%%%%%%%%%%%%%%%%%%%%%%%%%%%%%%%%%%%%%%%%%%%%%%%%%%%%%%%%%%%%%%%%%%%%%%%%%%%%%%%%%
%%%%%%%%%%%%%%%%%%%%%%%%%%%%%%%%%%%%%%%%%%%%%%%%%%%%%%%%%%%%%%%%%%%%%%%%%%%%%%%%%%%%%%%%%%%%
\section*{Data Availability Statement}

The data that support the findings of this study are available from the corresponding author upon reasonable request.

%%%%%%%%%%%%%%%%%%%%%%%%%%%%%%%%%%%%%%%%%%%%%%%%%%%%%%%%%%%%%%%%%%%%%%%%%%%%%%%%%%%%%%%%%%%%
%%%%%%%%%%%%%%%%%%%%%%%%%%%%%%%%%%%%%%%%%%%%%%%%%%%%%%%%%%%%%%%%%%%%%%%%%%%%%%%%%%%%%%%%%%%%
\appendix

\section{Appendixes}

\subsection{Slightly Detuned Resonators}
\label{sec:app:wL}

Let $\omega_1$ and $\omega_2$ be the channel frequencies of channel $1$ and $2$, respectively, and $\Delta\omega=\omega_1-\omega_2$. Only $\omega_1$ in channel $1$ is excited and we approximate the impedance of the coupled second resonator for small $\Delta\omega$. $R_\tu{s,2}$ is the real part of the resistance of channel $2$ at resonance.
The first order Taylor series expansion of the impedance of channel $2$ yields
\begin{align}\nonumber
    Z_2(\omega_2 + \Delta\omega) &= Z(\omega_2) + \Delta\omega \cdot \left. \frac{\text{d} Z(\omega)}{\text{d} \omega}\right|_{\omega_2} \\\nonumber
    &= R_\tu{s,2} + \Delta\omega\, j \left( L + \frac{1}{\omega_2^2 C} \right) \\\label{eq:appL}
    &= R_\tu{s,2} + j 2 \Delta\omega\, L \; . 
\end{align}
We insert $L=\tfrac{1}{\omega_2^2C}$ which is an expression of the resonance condition $\text{Im}(Z_2)=0$ at $\omega_2$ for a RLC resonant circuit. Here, $L$ is the sum of all inductances within the circuit. Consequently, we can rewrite \eqref{eq:appL} into
\begin{align}\nonumber
    Z_2(\omega_2 + \Delta\omega) &= R_\tu{s,2} \left( 1 + j2 \frac{\Delta\omega}{\omega_2}\frac{\omega_2 L}{R_\tu{s,2}} \right) \\
    &= R_\tu{s,2} \left( 1 + j2 \frac{\Delta\omega}{\omega_2} Q \right).
\end{align}
With a $\tfrac{\Delta\omega}{\omega_2}$ of $\tfrac{1}{75}$ and a $Q > 200$, we obtain a dominating imaginary part and the equation simplifies to  $Z_2(\omega_2 + \Delta\omega) \approx j2\Delta\omega L$.

%%%%%%%%%%%%%%%%%%%%%%%%%%%%%%%%%%%%%%%%%%%%%%%%%%%%%%%%%%%%%%%%%%%%%%%%%%%%%%%%%%%%%%%%%%%%
%%%%%%%%%%%%%%%%%%%%%%%%%%%%%%%%%%%%%%%%%%%%%%%%%%%%%%%%%%%%%%%%%%%%%%%%%%%%%%%%%%%%%%%%%%%%

\bibliography{MPILiterature_zotero.bib}

\end{document}